\documentclass{aa}  

\usepackage{graphicx}
\usepackage{txfonts}
\usepackage{hyperref}
\usepackage{natbib}
\begin{document}

\title{Onset of planet formation in the warm inner disk}
\subtitle{Colliding dust aggregates at high temperatures}
\titlerunning{Colliding Dust Aggregates at High Temperatures}

\author{Tunahan Demirci \and Corinna Krause \and Jens Teiser \and Gerhard Wurm}
\authorrunning{Demirci et al.}

\institute{University of Duisburg-Essen, Faculty of Physics, Lotharstr. 1-21, D-47057 Duisburg, Germany}

\date{Received 25 April 2019 / Accepted 26 July 2019}
 
\abstract
{}
{Collisional growth of dust occurs in all regions of protoplanetary disks with certain materials dominating between various condensation lines. The sticking properties of the prevalent dust species depend on the specific temperatures. The inner disk is the realm of silicates spanning a wide range of temperatures from room temperature up to sublimation beyond $1500\,\mathrm{K}$.}
{For the first time, we carried out laboratory collision experiments with hot levitated basalt dust aggregates of $1\, \rm mm$ in size. The aggregates are compact with a filling factor of $0.37 \pm 0.06$. The constituent grains have a wide size distribution that peaks at about $0.6\,\mu \mathrm{m}$. Temperatures in the experiments are varied between approximately $600\,\mathrm{K}$ and $1100\,\mathrm{K}$.}
{Collisions are slow with velocities between $0.002\,\mathrm{m}\,\mathrm{s}^{-1}$ and $0.15\,\mathrm{m}\,\mathrm{s}^{-1}$, i.e., relevant for protoplanetary disks. Aside from variations of the coefficients of restitution due to varying collision velocities, the experiments show low sticking probability below $900\,\mathrm{K}$ and an increasing sticking probability starting at $900\,\mathrm{K}$.}
{This implies that dust can grow to larger size in hot regions, which might change planet formation. One scenario is an enhanced probability for local planetesimal formation. Another scenario is a reduction of planetesimal formation as larger grains are more readily removed as a consequence of radial drift. However, the increased growth at high temperatures likely changes planetesimal formation one way or the other.}


\maketitle

\section{Introduction}

It is common knowledge by now that planet formation regularly takes place in protoplanetary disks surrounding young stars \citep{Keppler2018}. It is also undisputed that planet formation starts with dust grains that collide, stick together, and grow \citep{Blum1998}. How far they grow and how they evolve to planetesimals and planets, eventually, is a matter of debate. There is a bouncing barrier at millimeter-size \citep{Zsom2010, Guettler2010}.
At this barrier, compact dust aggregates typically collide at $\mathrm{mm}\, \mathrm{s}^{-1}$ to $\mathrm{cm}\, \mathrm{s}^{-1}$ but no longer stick and rather bounce off each other. Lucky winners have been proposed to lead to further growth in exceptionally low speed collisions \citep{Windmark2012, Drazkowska2013}. However, experiments show that the connections formed in sticking events are very weak and any subsequent faster collision disassemble such aggregates again \citep{Kelling2014, Kruss2016}. If reservoirs of small grains are provided, the aggregates might grow further by sweeping them up \citep{Kruss2017}, but this would require a fine-tuned source of grains. Dust released at the snowline might be a source \citep{Saito2011, Aumatell2011}, although this would act very locally. If aggregates grew only slightly larger and collided only slightly faster they would encounter fragmentation as a next potential barrier to growth \citep{Beitz2011, Deckers2014, Birnstiel2016}.
Growth can still proceed in collisions between different sized aggregates, even at high speeds of tens of $\mathrm{m}\, \mathrm{s}^{-1}$ \citep{Wurm2005, Teiser2009, Windmark2012}. Again, there is a requirement in this case. A few larger seeds are needed to start with and the process is rather slow.

Mechanisms that might circumvent this troublesome climb in size have been proposed. A suite of instabilities might act to concentrate solids \citep{Klahr2018}. If the surface mass density of solids is large enough, gravity is able to lead to a direct planetesimal formation.
The underlying motion of grains and capabilities of the concentration mechanisms strongly depend on the Stokes number, which is the ratio of the gas-grain coupling time to the orbital time. If grains are too small, they perfectly couple to the gas and concentration does not work. If grains are too large, they are essentially decoupled from the gas and concentration does not work either.
Minimum sizes required in the inner disk region are on the order of centimeters \citep{Yang2017}. This is somewhat beyond the bouncing and fragmentation sizes currently. Details of the respective simulations are beyond the scope of this paper. It should be noted that reaching certain aggregate sizes by collisional growth does not mean that the formation of planets is warranted, as there are also erosive mechanisms during planet formation that can destroy larger objects again. For example, \citet{Demirci2019} recently showed that planetesimals formed that way might be unstable to wind erosion in protoplanetary disks. Anyway, what is important to note is that it matters strongly for these concentration mechanisms to what size  exactly aggregates can grow in collisions before encountering bouncing and fragmentation. The larger the particles become, the more likely growth yields to further evolution. Small changes in sticking properties might change that picture significantly.

The importance of sticking has, for example, been attributed to water ice. It is often taken for granted that water ice sticks better. The surface energy often considered is on the order of $\gamma = 0.1\,\mathrm{J}\,\mathrm{m}^{-2}$ \citep{Dominik1997, Gundlach2011, Aumatell2013}. In comparison values for silicate surface energies taken have been an order of magnitude lower \citep{Dominik1997, Heim1999}. Assuming this and a $\mu \rm{m}$ grain size \citet{Okuzumi2012} and \citet{Kataoka2013} modeled the growth of gigantic fluffy ice aggregates that pass bouncing, fragmentation, and a drift barrier. The latter is the inward drift from the sub-Keplerian motion of the gas that drains the dust reservoir \citep{Weidenschilling1977, Birnstiel2010}.

In any case, it has recently been shown that this simple picture of non-sticky silicates and sticky water ice grains does not hold as a general idea. In fact, the surface energy of water decreases by almost two orders of magnitude going from 273 K temperature to $180\,\rm K$ \citep{Musiolik2019}. This latter temperature, and even lower temperatures, are relevant for most parts of protoplanetary disks in which water ice dominates.
On the other side, the surface energy of silicates at the dry conditions of protoplanetary disks is an order of magnitude higher than previously assumed \citep{Kimura2015, Steinpilz2019}.
So water ice is not stickier at low temperatures \citep{Gundlach2018}. Only close to the water snowline does water ice rule sticking.

With these facts on water ice in mind, it is a logical next step to ask what influence the temperature has on silicate sticking properties. This is especially important in the inner part of protoplanetary disks, where temperatures finally reach the sublimation limit.
There are only few works so far on collision properties considering high temperatures, starting with tensile strength measurements of tempered dust by \citet{deBeule2017} to collisions of tempered aggregates by \citet{Demirci2017} to first hot collisions of chondrule size grains by \citet{Bogdan2019}. These works are, as yet, inconclusive on the effect of particle growth but indicate that strong changes in sticking properties occur beyond 900 to $1000\,\rm K$.
In this work, we report on the first experiments of slow dust aggregate collisions at temperatures up to about $1100\,\mathrm{K}$.

\section{Experiment}

\begin{figure}
        \includegraphics[width=\columnwidth]{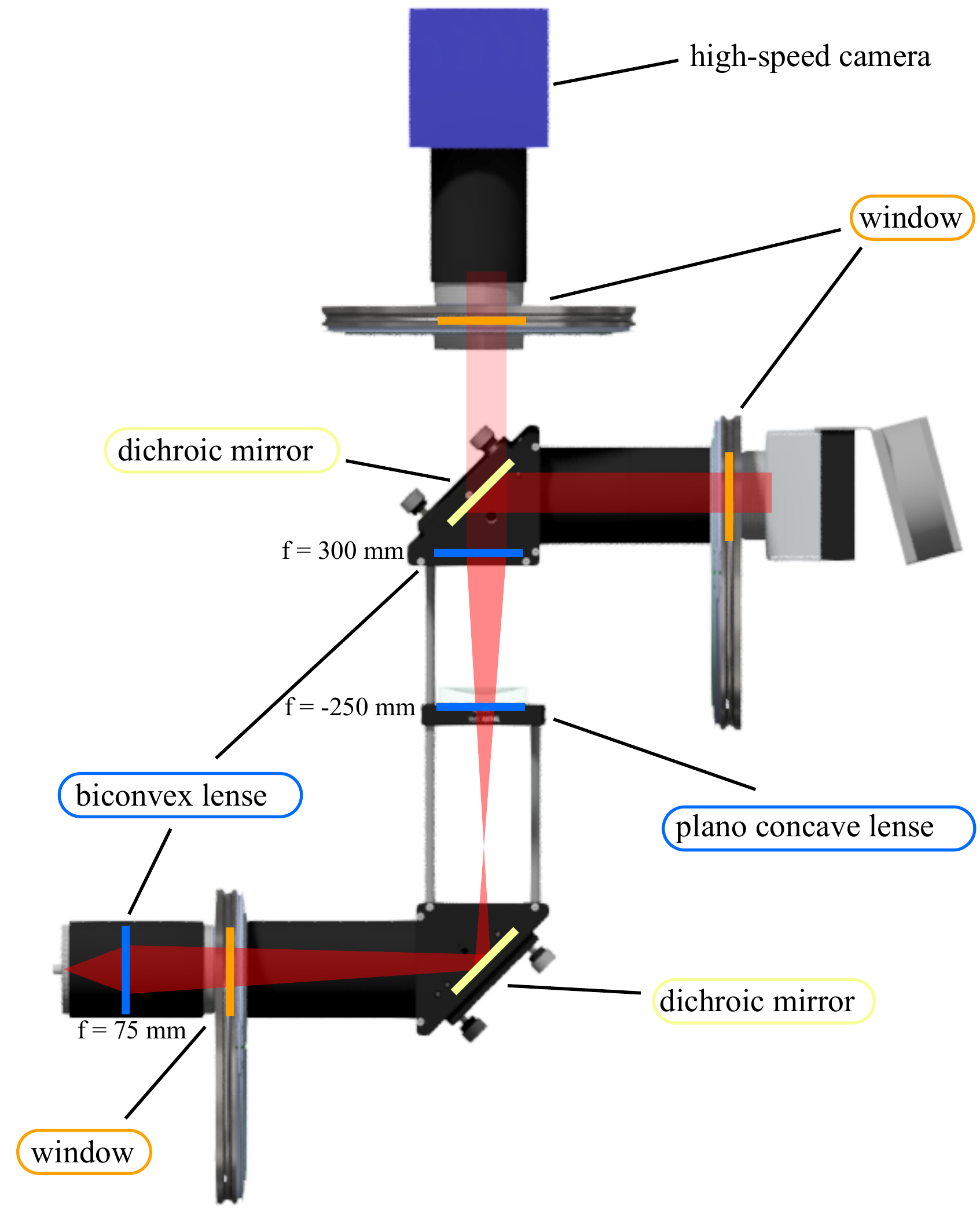}
        \caption{\label{fig.setup} Basic sketch of the experimental setup. The basalt aggregates are placed on the plano concave lense which serves as an experimental platform.}
\end{figure}

Collision experiments require free particles. Most of the experiment is therefore a mechanism to levitate dust aggregates. We use the principle of levitating aggregates at low pressure by a Knudsen compressor
\citep{Kelling2009}. This has been used in various studies before \citep{Kelling2014, Kruss2016, Kruss2017, Demirci2017, Kruss2018}. Also ice aggregates have been levitated by this mechanism \citep{Aumatell2011}. Details of the levitation can be found in these references. The basic idea is that a temperature gradient at low pressure leads to thermal creep. For particle aggregates on a hot surface this gas flow is directed downward and aggregates create their own air cushion to float upon. However, earlier experiments could not heat the particles much higher than $800\,\rm K$ for technical reasons.  We therefore developed a new experimental setup to allow heating well beyond $1000\,\rm K$ as seen in fig. \ref{fig.setup}.
We used an infrared laser beam to avoid heating up the entire chamber. The wavelength of the near-infrared laser is $934.6\,\rm nm$. The maximum power is $45\, \rm W$. With a beam cross section of $1.45\, \mathrm{cm}^2$ the maximum laser intensity is about $311\,\mathrm{kW}\,\mathrm{m}^{-2}$. The laser beam selectively heats the dust aggregates but the transparent surface on which the sample is placed stays cool. The laser is reflected at a dichroic mirror and leaves the chamber again and the energy is placed in a beam dump. The small fraction that is not reflected serves as background illumination for the grains.

\begin{figure}
	\begin{center}
		\includegraphics[width=0.8\columnwidth]{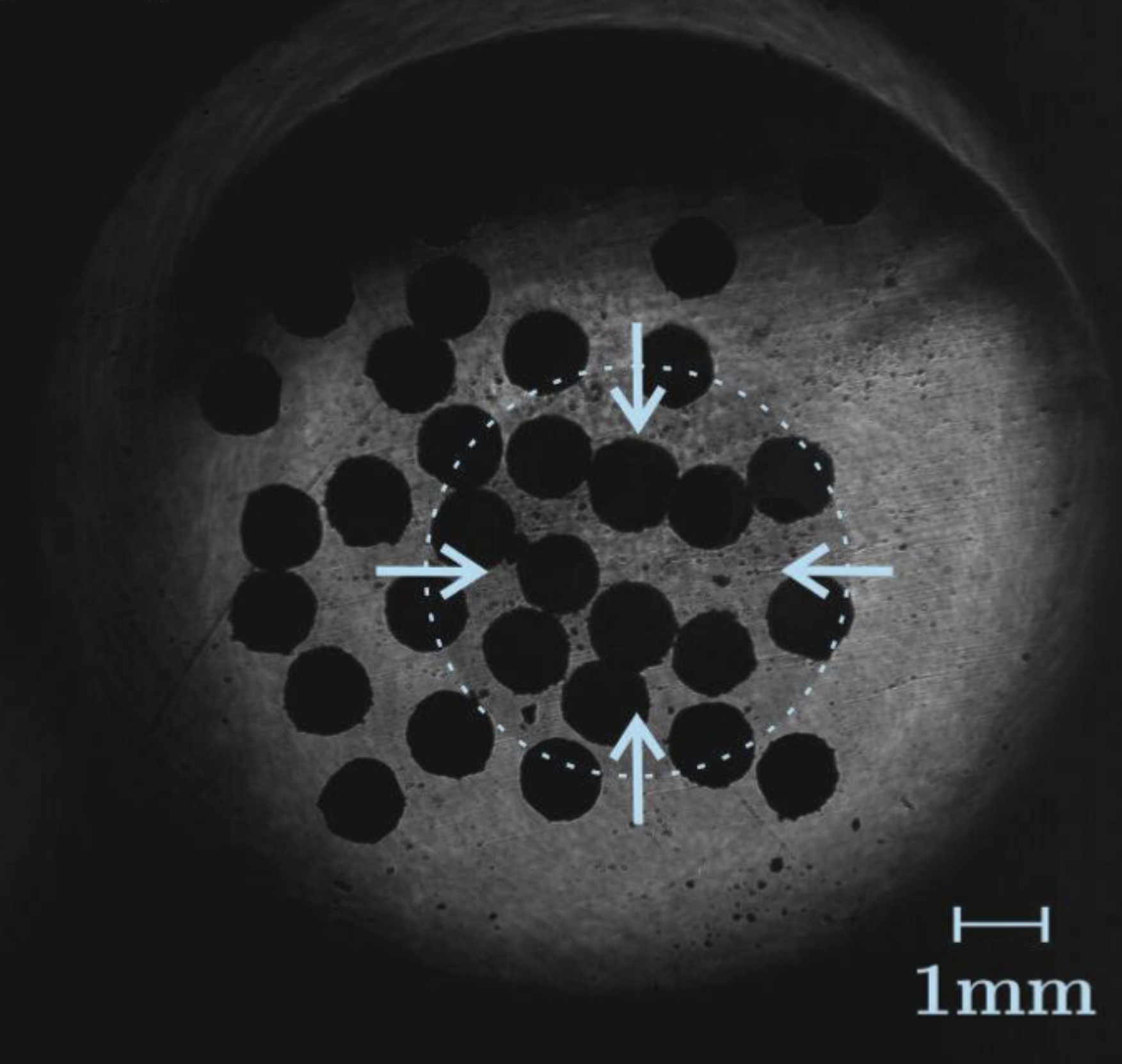}	
	\end{center}
	
	\caption{\label{fig.profile} Top view of the experiment platform with the levitated basalt aggregates. The gold coated center is highlighted in the picture. The collision dynamics of the basalt aggregates are recorded with a frequency of 1000 frames per second.}
\end{figure}

\begin{figure}
        \begin{center}
                \includegraphics[width=0.8\columnwidth]{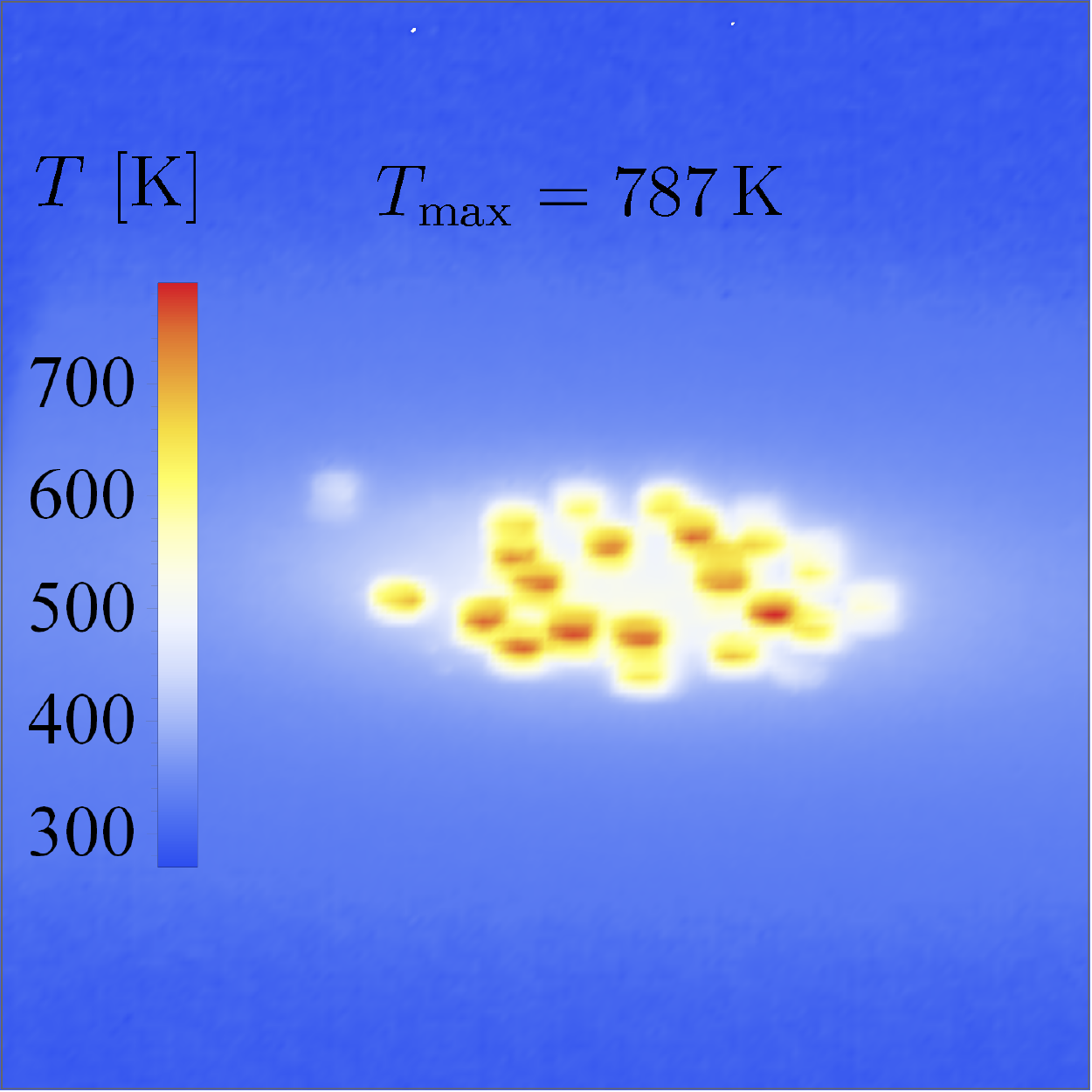}
        \end{center}
        
        \caption{\label{fig.thermogram} Thermal image of the experimental platform with the basalt aggregates. The laser illuminates the aggregates and heats them while the platform stays cool. The maximum temperature in this picture is $787\,\rm K$.}
\end{figure}

Particles levitating are free to move in any direction. Eventually, these particles leave the illuminated part and settle on the surface again. To delay this process,
the beam profile is shaped with an intensity minimum at the center, which can be seen in fig. \ref{fig.profile}. Particles therefore also get a lateral temperature gradient, which effectively keeps particles trapped in the center. The local intensity minimum at the center of the experiment platform is generated by coating the center of the sample holder with a thin gold layer. For a given ambient pressure and given laser intensity (controlled with the driver current) the temperature of the dust aggregates was measured with an infrared camera as seen in fig. \ref{fig.thermogram}.

\begin{figure}
        \includegraphics[width=\columnwidth]{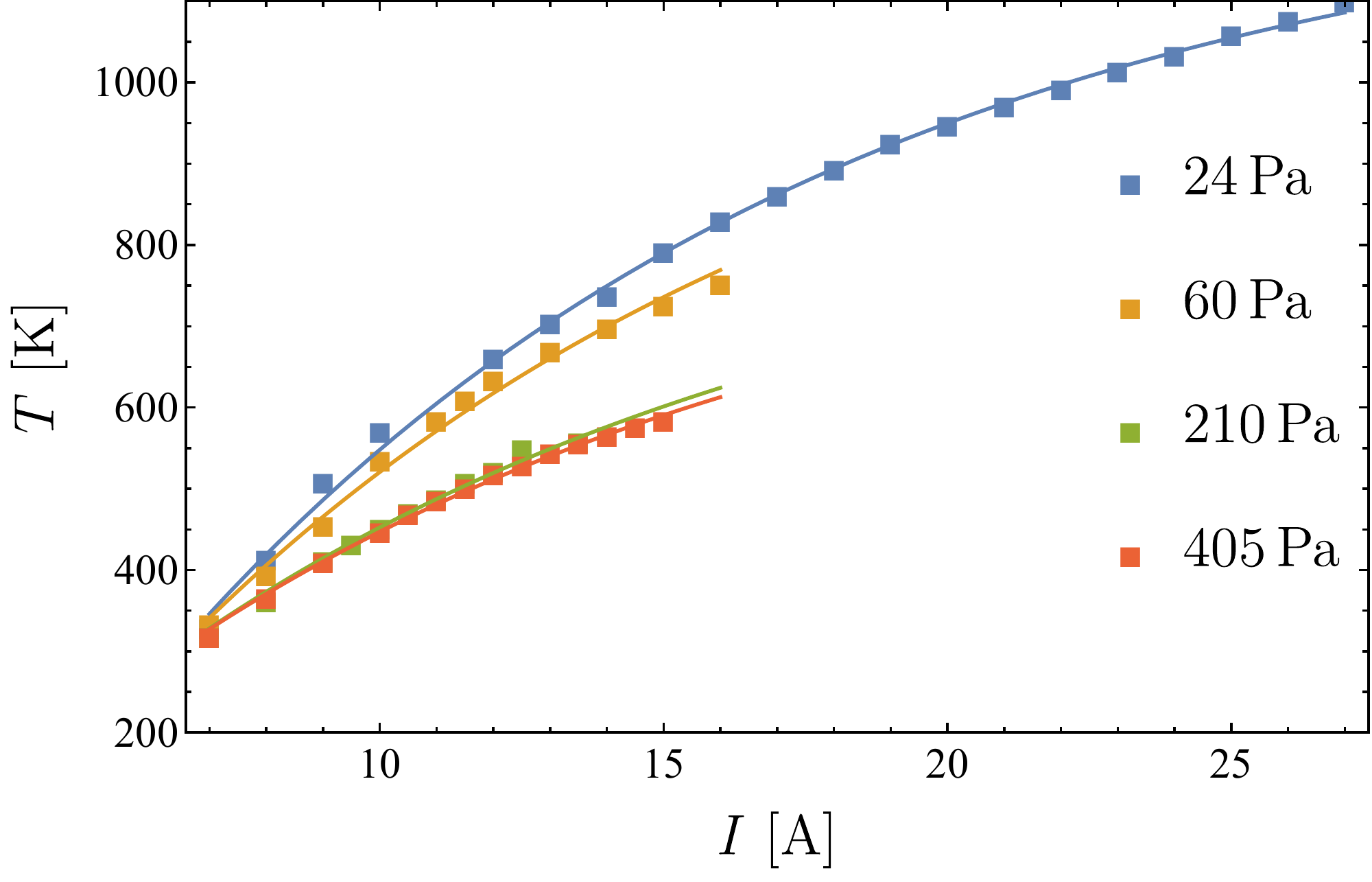}
        \caption{\label{fig.calibration}Temperature calibration measurements for the basalt aggregates at ambient pressures $p \in [24,405]\,\rm Pa$ and laser driver current $I \in [7,27]\,\rm A$. The temperature is measured with a thermal camera assuming an emissivity $\epsilon=0.72$ for basalt \citep{Golden2013}. The function ${T(I,p)=T_0+T_\infty(p_1) (\frac{p}{p_1})^{-\alpha} \left(1-\exp\left(-\frac{I-I_\Delta}{I_0}\right) \right)}$, with fit parameters  $T_0$, $T_\infty(p_1)$, $\alpha$, $I_\Delta$ and $I_0$, is fitted to the data and is used to determine the temperatures for the experimental runs.}
\end{figure}

Calibration curves were recorded to calculate the temperature of aggregates for a given pressure and intensity combination as seen in fig. \ref{fig.calibration}. We neglect the influence of the beam shaping upon the local temperature. We also do not consider but note that owing to the levitation principle there is a temperature difference between top and bottom of the aggregate of about $20\,\mathrm{K}$ \citep{Kelling2014}.

\begin{figure}
        \includegraphics[width=\columnwidth]{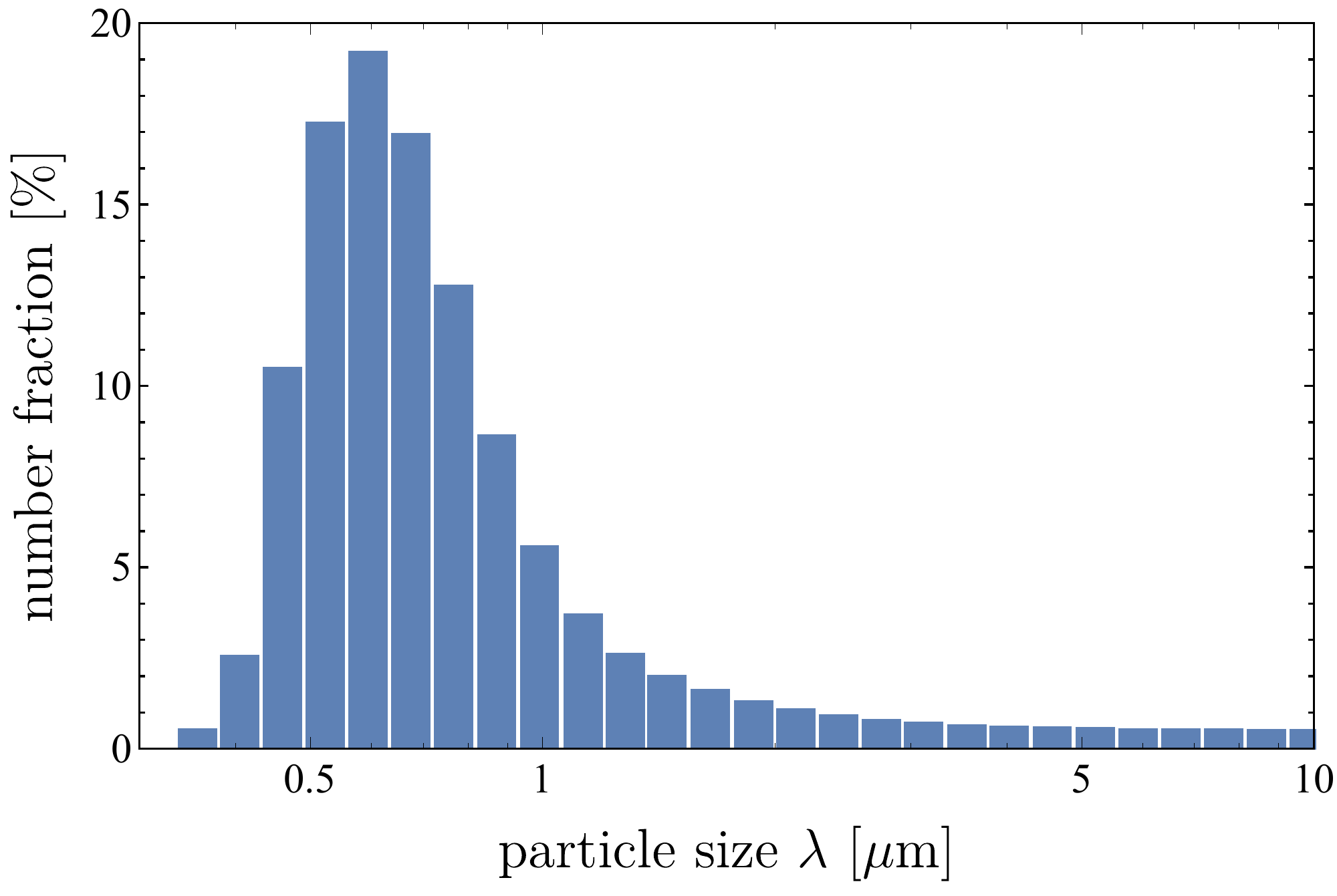}
        \caption{\label{fig.sizedistribution} Number size distribution of the initial basalt sample. The distribution was measured by a commercial instrument based on light scattering (Malvern Mastersizer 3000).}
\end{figure}

A high-speed camera on top records the motion of the particles. This motion is tracked afterward, resulting in relative (collision) velocities $\vec{v}_\mathrm{c}$ before, the relative (rebound) velocities $\vec{v}_\mathrm{r}$ after a collision, and the connection vector $\vec{r}$ between two aggregates. Based on these quantities we define a coefficient of restitution
\begin{equation}
k = \frac{v_\mathrm{r}}{v_\mathrm{c}}
\end{equation}
and the impact parameter
\begin{equation}
b = \frac{\lVert \vec{v}_\mathrm{c} \times \vec{r} \rVert}{v_\mathrm{c} r}.
\end{equation}
In this equation $v_c$ and $r$ are the norm of the corresponding vectors. As dust is leaving the illuminated spot on short timescales in spite of the beam shaping, we cannot observe a self-consistent growth of aggregates and only observe individual collisions.
As samples we therefore used aggregates, which we artificially produce before by pressing basalt powder (approx. $0.6\,\mu \mathrm{m}$ in size, see fig. \ref{fig.sizedistribution}) in a mold of $1\,\mathrm{mm}$ diameter and $0.2\,\mathrm{mm}$ height. They have a filling factor of $0.37 \pm 0.06$.

\section{Results}

\begin{figure}
        \includegraphics[width=\columnwidth]{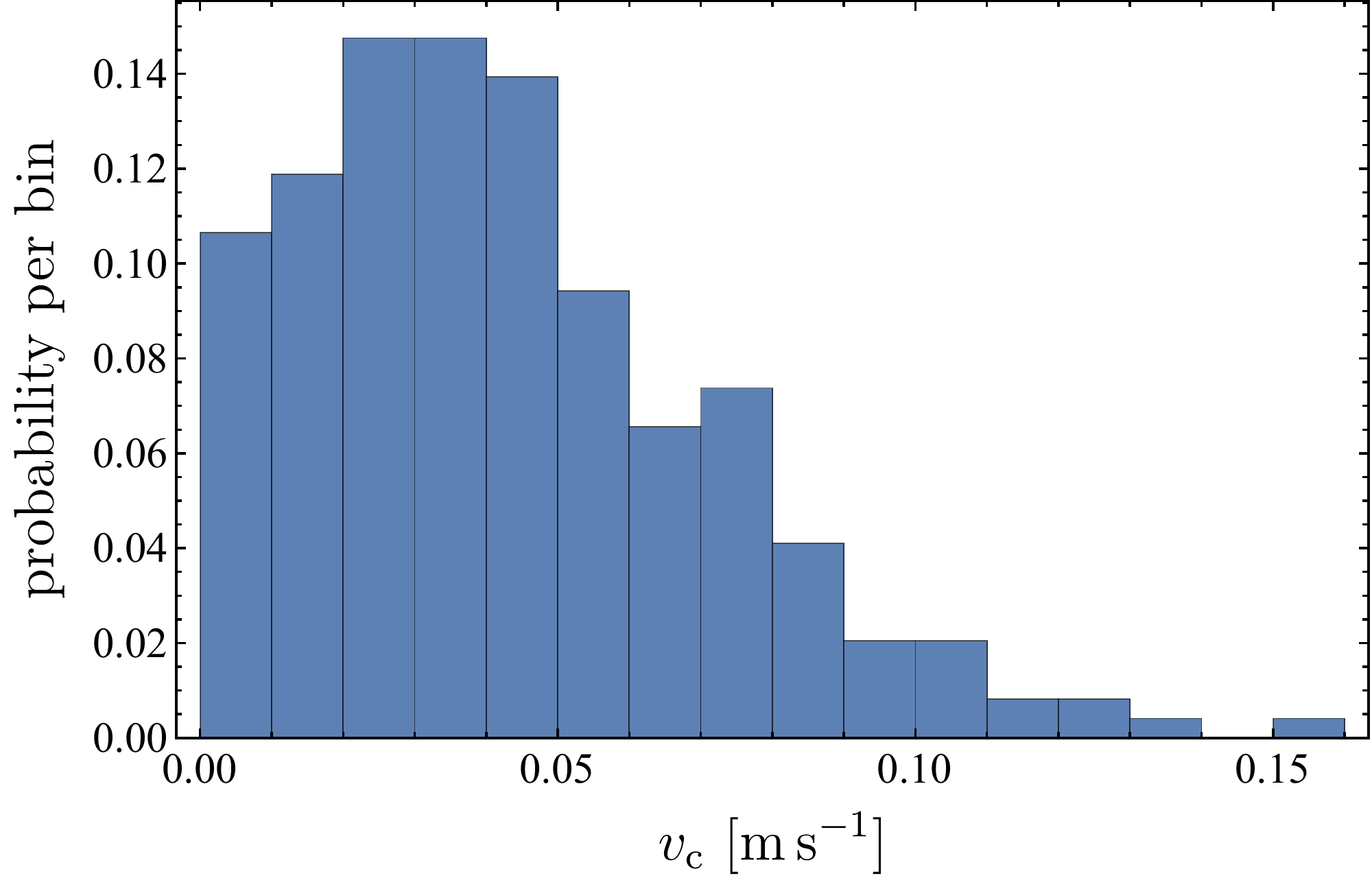}
        \caption{\label{fig.velocities} Histogram of the collision velocities $v_\mathrm{c}$. The mean collision velocity is $v_\mathrm{c}^\mathrm{mean}=0.04\,\mathrm{m}\,\mathrm{s}^{-1}$ and the maximum collision velocity is $v_\mathrm{c}^\mathrm{max}=0.15\,\mathrm{m}\,\mathrm{s}^{-1}$. In total we analyzed 244 collisions between two individual aggregates.}
\end{figure}

In total we observed 244 collisions between two individual aggregates. Outcomes of collisions were only rebound and sticking.
Typical collision velocities observed are a few $\mathrm{cm}\, \mathrm{s}^{-1}$. A distribution of all collision velocities
is shown in fig. \ref{fig.velocities}.

\begin{figure}
        \includegraphics[width=\columnwidth]{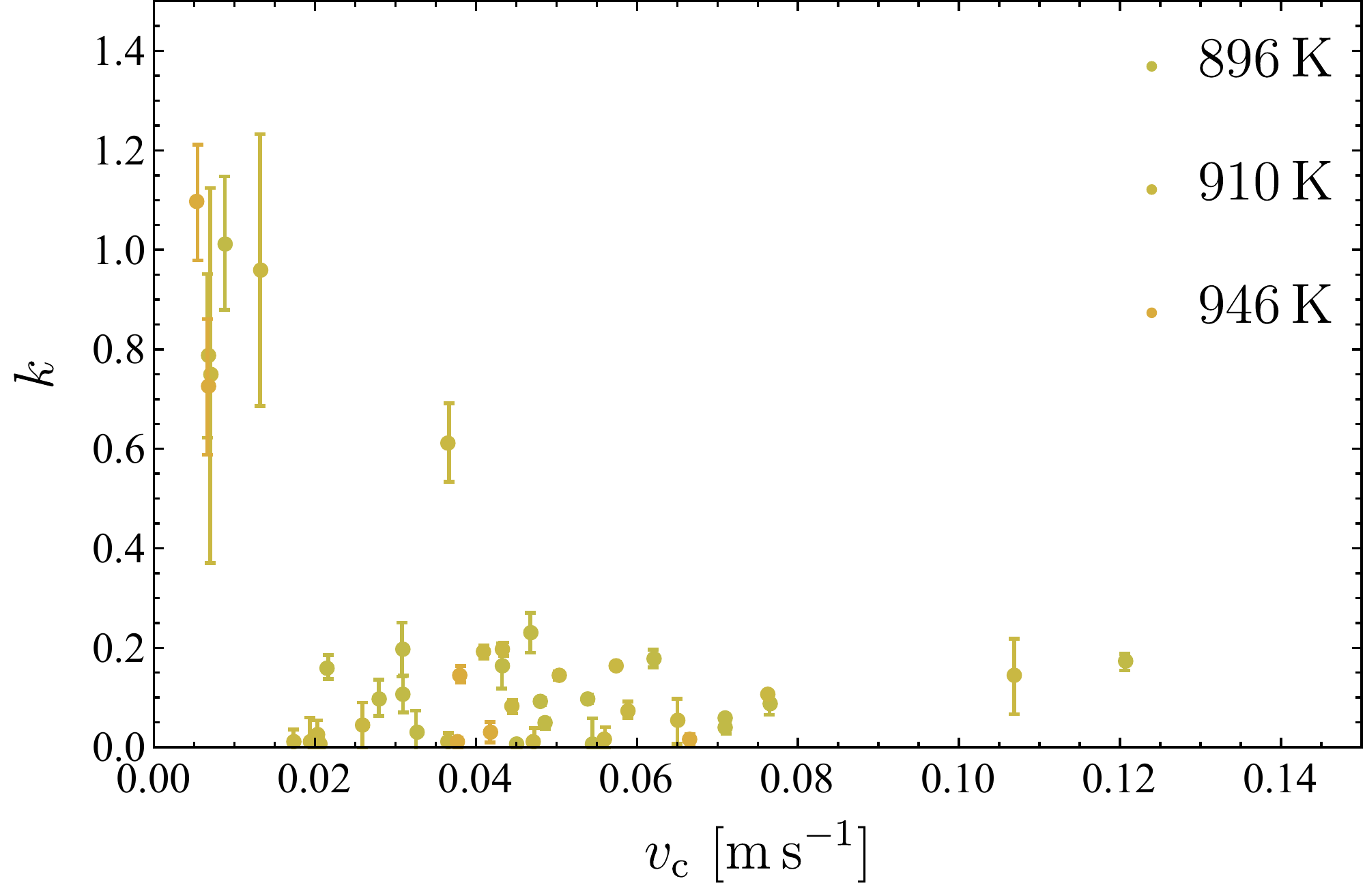}
        \caption{\label{fig.musiolik}Typical dependence of the coefficient of restitution on the collision velocity for the temperature interval  $T=  896\, - \, 946\, \rm K$. All data are separated in a high-$k$ region at low velocities and a low-$k$ region beyond about 2 cm/s, which also includes sticking events with $k=0$.}
\end{figure}

We considered a dependence of the coefficient of restitution on the impact parameter but did not see any clear dependence. The Pearson correlation coefficient between k and b is only 0.35, so we do not evaluate this further.
The coefficient of restitution clearly depends on the collision velocities though. Fig. \ref{fig.musiolik} shows a typical example adding data at three neighboring temperatures. While we compared the data to typical models of restitution coefficients, i.e., by \citet{Musiolik2016}, the data could not be fitted well and no further insights could be gained. Instead, it can be seen that the data is essentially bimodal. At low collision velocities the coefficient of restitution is high. At velocities beyond 2 cm/s the coefficient of restitution is low including sticking events. The average collision velocities $\bar{v}_\mathrm{c}$ do not depend on the temperature. Also, the standard deviation, $\sigma_{v_\mathrm{c}}$, is very large. Therefore the velocity ranges are very similar as seen in table \ref{tab:velocities}.

\begin{table}
\begin{center}
\caption{\label{tab:velocities} Average collision velocities for multiple temperature ranges with standard deviations. To exclude collision events near the sticking velocity only collisions with velocities greater than $0.02\,\mathrm{m}\,\mathrm{s}^{-1}$ are taken into account.}
\begin{tabular}{c|c|c|c|c}
$T$&$\bar{v}_\mathrm{c}$&$\sigma_{v_\mathrm{c}}$&$v_\mathrm{c,min}$&$v_\mathrm{c,max}$\\
$[\mathrm{K}]$&$[\mathrm{cm}\,\mathrm{s}^{-1}]$&$[\mathrm{cm}\,\mathrm{s}^{-1}]$&$[\mathrm{cm}\,\mathrm{s}^{-1}]$&$[\mathrm{cm}\,\mathrm{s}^{-1}]$\\
\hline
&&&&\\
596 -- 690&4.8&2.0&2.1&9.8\\
740 -- 850&5.7&2.9&2.1&15.2\\
896 -- 946&5.0&2.1&2.0&12.1\\
1040 -- 1085&5.6&3.0&2.4&12.2\\
\end{tabular}
\end{center}
\end{table}

\begin{figure}
        \begin{center}
                \includegraphics[width=0.7\columnwidth]{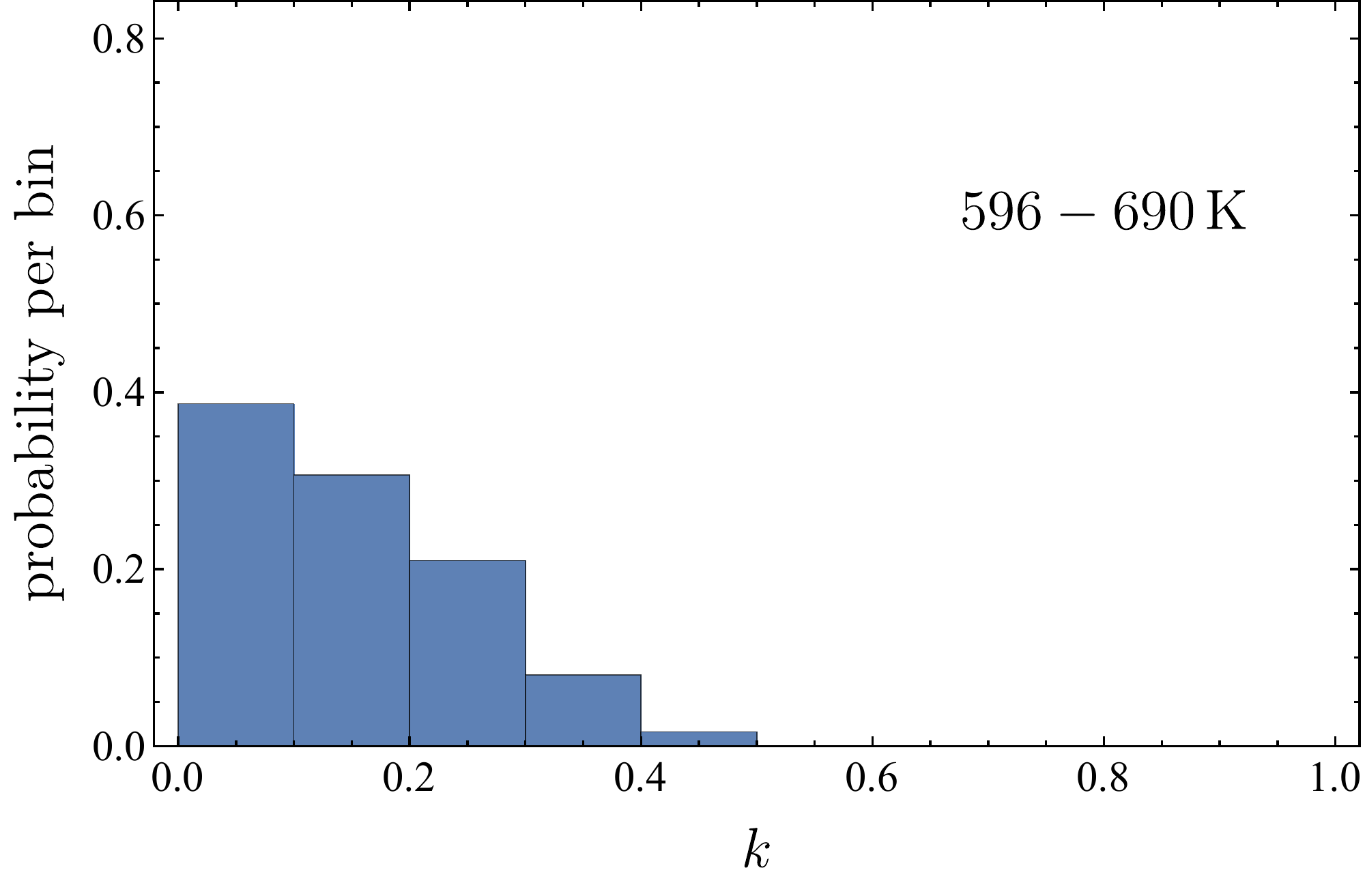}
                \includegraphics[width=0.7\columnwidth]{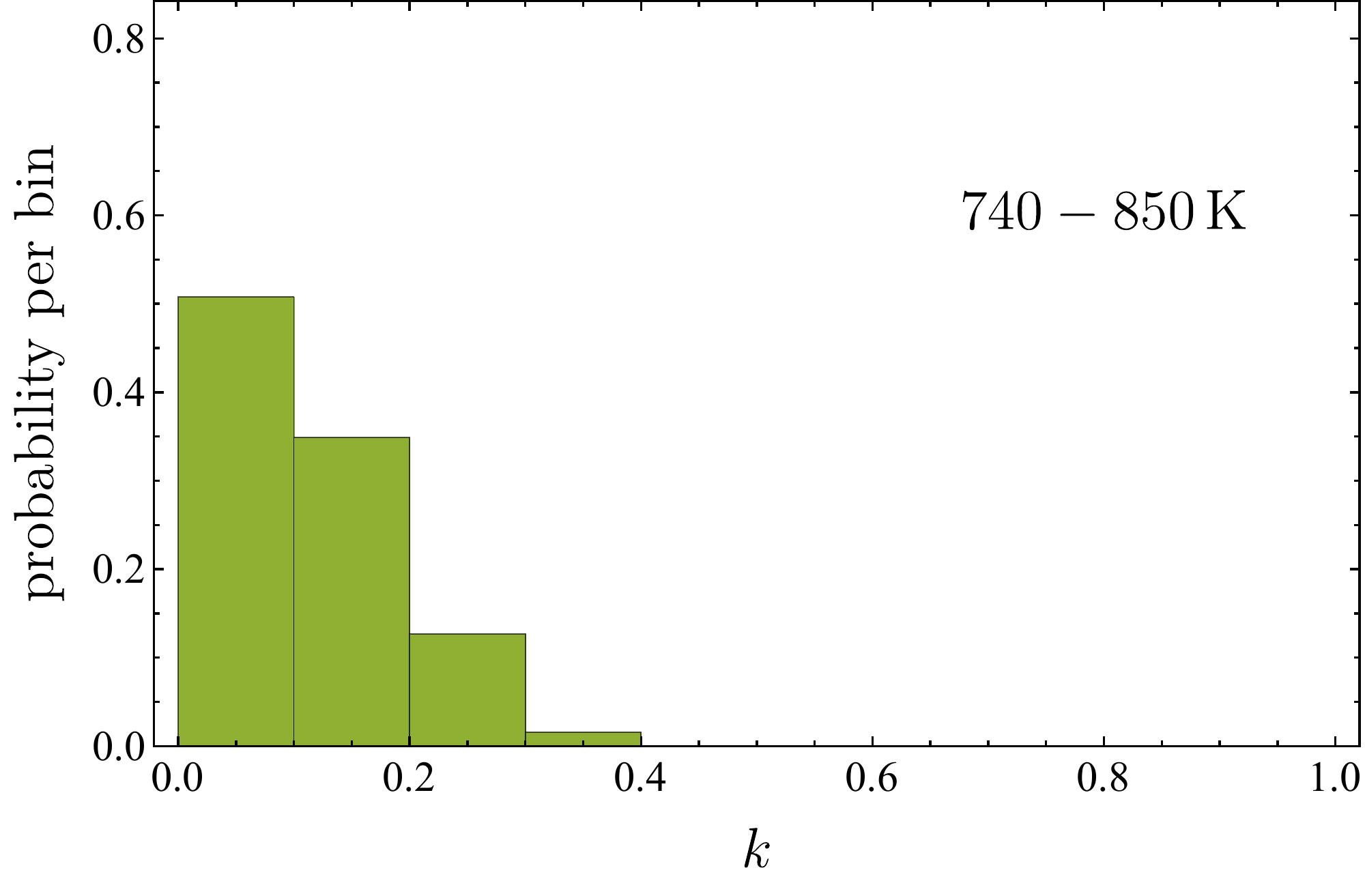}
                \includegraphics[width=0.7\columnwidth]{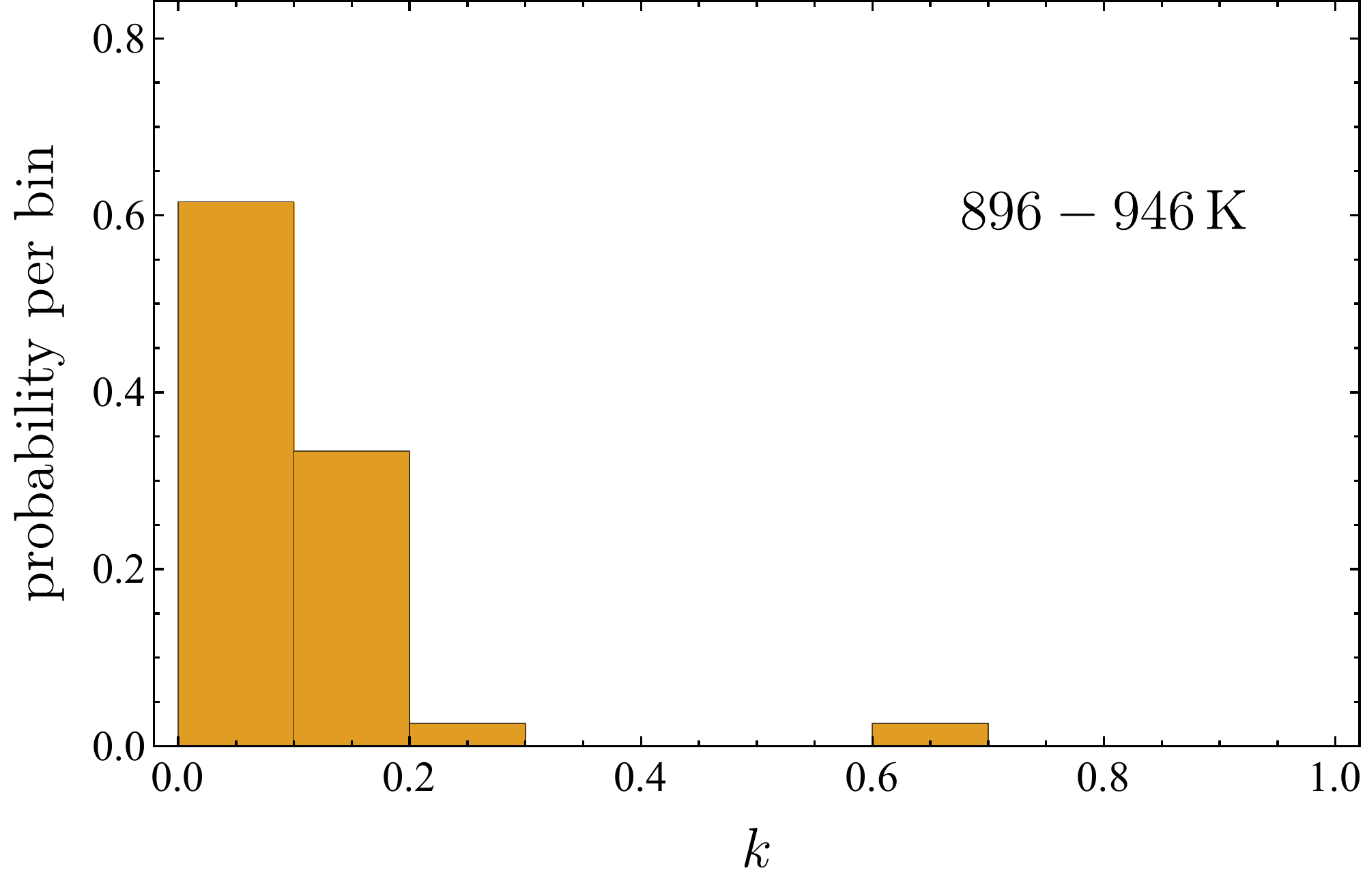}
                \includegraphics[width=0.7\columnwidth]{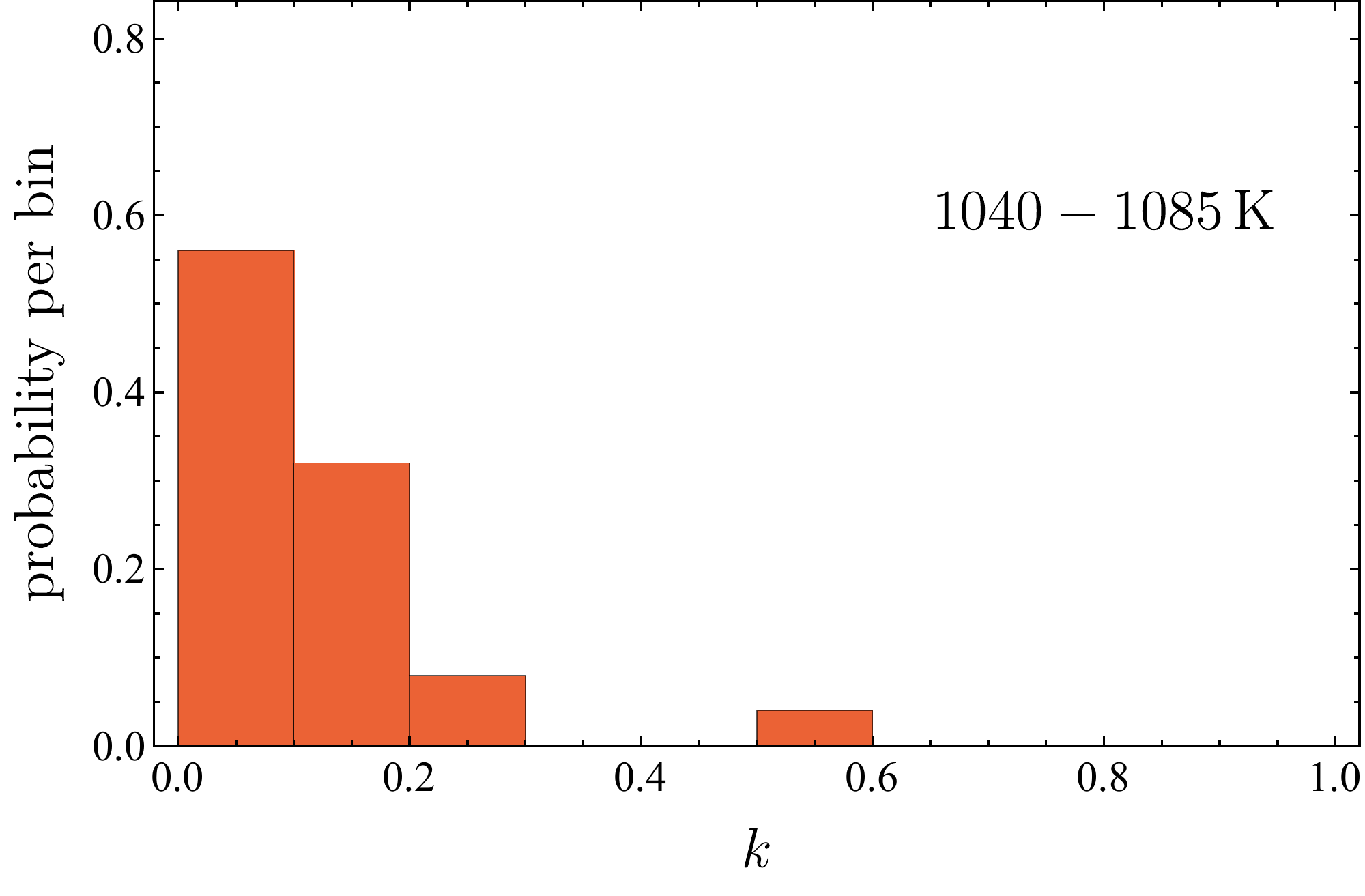}
        \end{center}
        \caption{\label{fig.histograms}Histograms of the coefficient of restitution $k$ for the different temperature intervals. To exclude collision events near the sticking velocity only collisions with velocities greater than $0.02\,\mathrm{m}\,\mathrm{s}^{-1}$ are taken into account.}
\end{figure}

Based on this overlap in collision velocities we can compare the coefficients of restitution for all events with velocities $v_\mathrm{c}\geq 0.02\,\mathrm{m}\,\mathrm{s}^{-1}$.
The distributions of $k$ for different temperatures are seen in fig. \ref{fig.histograms}. Up to $946\,\mathrm{K}$ the coefficient of restitution clearly decreases with temperature. Above that temperature no more significant difference is seen in the distribution.

\begin{figure}
        \includegraphics[width=\columnwidth]{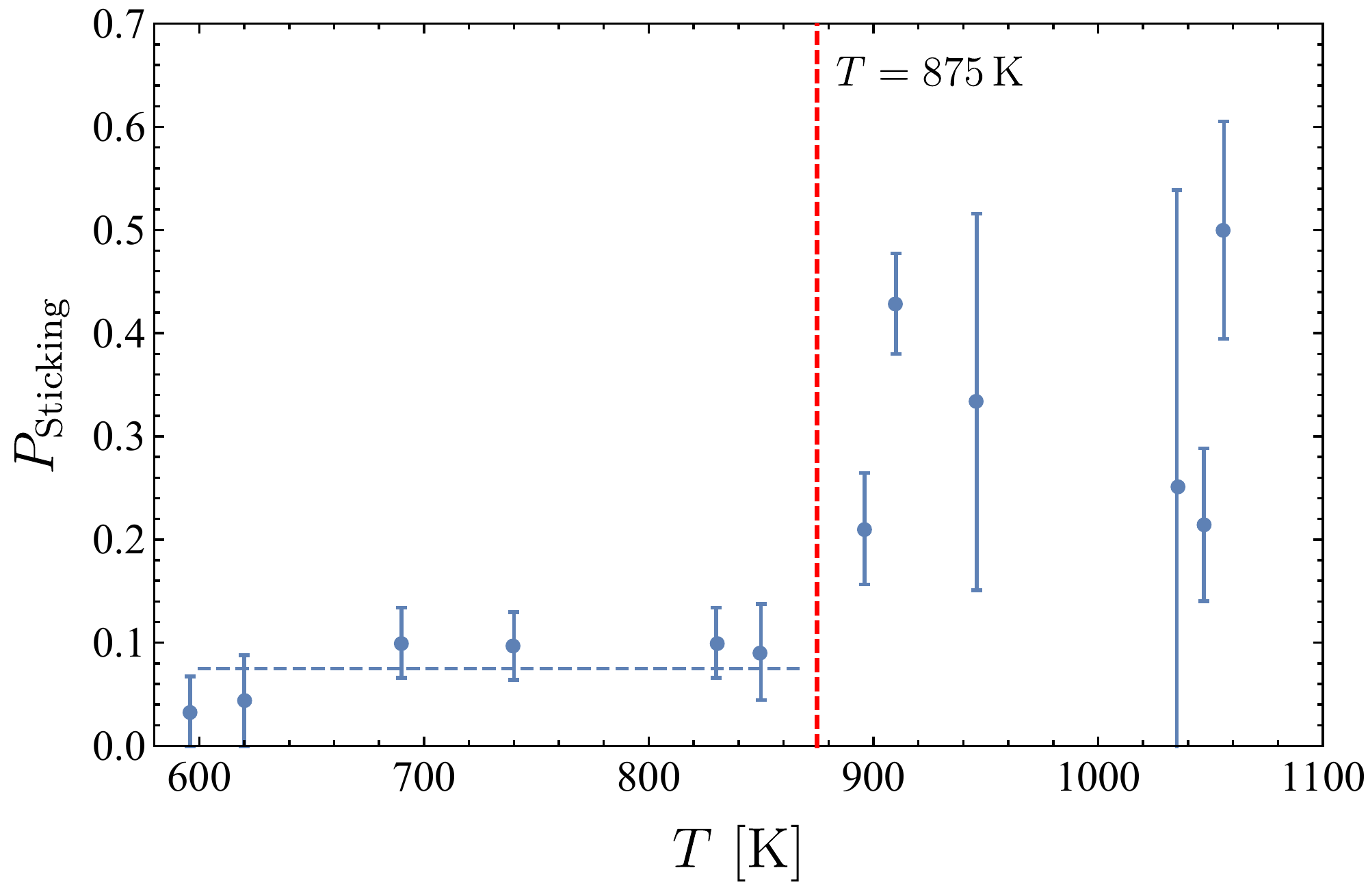}
        \caption{\label{fig.stickpro}Sticking probability $P_\mathrm{Sticking}$ in dependence of the temperature $T$. For $T>900\,\mathrm{K}$ the collisions of the basalt aggregates show an increase in the sticking probability. The temperatures are binned in an interval of $5\,\rm K$ and only temperature intervals with more than 4 collisions are taken into account. The $875\,\rm K$ line is plotted in red. Indeed, collisions beyond this line result in more sticking.}
\end{figure}

The sticking events, which we observed, occur at higher collision velocities than predicted by simple models for sticking collisions. In detail, dust aggregates can stick over a wide range of velocities. There is no clear separation in sticking and rebound. We therefore use a sticking probability to characterize the temperature dependence. A value of $1$ is attributed to all sticking events and a value of $0$ is attributed to rebound. For each temperature interval all values are summed up over all collision velocities and normalized by the total number of collisions. The results are shown in fig. \ref{fig.stickpro}. There is a clear dependence of the sticking probability on the temperature. It has to be noted that the coefficient of restitution $k$ obviously depends on the collision velocity $v_\mathrm{c}$. Therefore, to compare the temperature dependent sticking probability the velocities for the different temperature intervals must be comparable. We note that this is not exactly the case in our data.

\section{Discussion}

The measurements do not directly show that dust grains can grow to larger aggregates in hot environments.
However, they show that the stickiness increases significantly and in a self-consistent growth it is likely that aggregates at the bouncing barrier will be larger. We cannot quantify yet how much larger that will be,
but there will be a size difference.
How does this compare to the existing experiments on tempered dust or hot collisions?

On one side, earlier experiments by \citet{deBeule2017} show a peak in tensile strength for samples of JSC Mars 1a dust tempered at $1000\,\rm K$ in air. The tensile strength then decreased again to larger temperatures of tempering. Their data is consistent with a change in sticking properties close to $1000\,\rm K$. At higher temperatures sticking does not work well. This is  consistent with results from \citet{Demirci2017} in which a self-consistent evolution into the bouncing barrier was studied. They showed that basalt dust tempered beyond $1000\,\rm K$ grows to a smaller maximum size than basalt tempered at lower temperatures.

Real collision experiments with hot particles were carried out by \citet{Bogdan2019}. They used compact $1\, \rm mm$ basalt spheres and observed that the coefficient of restitution decreases or in other words the sticking forces increase starting at about $900\,\rm K$. This is consistent with the results reported in this paper. Using dust, the coefficient of restitution is generally lower and sticking can result more readily. Otherwise, the increased sticking probability also starts at about $900\,\rm K$ for basaltic dust.

At first sight, the experiments by \citet{Demirci2017} seem to be in contradiction to the experiments by \citet{Bogdan2019}. It has to be kept in mind though that in the first experiments the sample was measured cooler than the temperature tempered at while in the last experiments samples were really hot during the experiments. This way, the effect of changes in mineral content or grain size onto stickiness are traced in the first experiments referred to. This is different from carrying out the experiments hot, which changes, for example,  the viscosity, plasticity, or surface energies of the material.

As the later experiments, i.e., the experiments reported in this work, were carried out at high temperature they include the effects of changing static properties of the sample, for example, composition and grain size but also the dynamic properties of increased sticking, for example, due to plasticity at high temperatures. Obviously, the increase in sticking properties due to high temperatures wins over reducing sticking effects of other particle properties.

With the knowledge of the mineralogical composition of the basalt powder\footnote{example basalt composition (Kremer Pigmente): $50\,\%\, \mathrm{SiO}_2$, $2.5\,\%\, \mathrm{TiO}_2$, $12.1\,\%\, \mathrm{Al}_2 \mathrm{O}_3$, $2.1\,\%\, \mathrm{Fe}_2 \mathrm{O}_3$, $10\,\%\, \mathrm{Fe}\mathrm{O}$, $0.16\,\%\, \mathrm{Mn} \mathrm{O}$, $9.6\,\%\, \mathrm{Mg}\mathrm{O}$} the temperature dependent viscosity can be calculated with the model of \citet{Giordano2008}, i.e.,
\begin{equation}
\eta_\mathrm{Basalt} (T)=2.3 \times 10^{-3} \exp \left(\frac{6384.4\,\mathrm{K}}{T-573.1\,\mathrm{K}}\right) \mathrm{Pa}\,\mathrm{s}.
\end{equation}
This equation describes the viscosity of basalt for temperatures near $1000\,K$. \citet{Hubbard2015} proposed that collisions between two particles behave sticky, if the viscosity fall below a critical value $\eta_\mathrm{s}=t_\mathrm{M} \mu$. In this case, $t_\mathrm{M}$ is the Maxwellian relaxation time of the material and $\mu$  the shear modulus. With $\mu=4 \times 10^{4}\,\mathrm{Pa}$ for chondrites and for typical collision timescales of $\Delta t= 10^{-4}\, \rm s$ \citep{Kelling2014} the viscosity of basalt is equal to the critical value at a temperature of $875\,\rm K$. So collisions of basalt aggregates above $875\,\rm K$ should lead to more sticking cases. Indeed, we see in our data that the probability of sticking collisions is higher above this temperature than below this temperature (see fig. \ref{fig.stickpro}). This behavior was also reported by \citet{Bogdan2019}. 

Collisions at a still higher temperature were also observed but due to too efficient levitation, aggregates, for example, flipped over, preventing a classification of these collisions. The experimental setup has the ability to study collisions up to sublimation temperatures and we plan to optimize the experiments to do this.

\section{Application to planet formation}

We cannot give an ultimate factor by which aggregate sizes will increase if dust grows at hot conditions. Obviously, dust can grow to larger sizes in the inner part of protoplanetary disks outside of the sublimation line of silicates. From our data we suggest that the region between the sublimation region and the $900\, \rm K$ line will stand out. \citet{Demirci2017} argued that $1000\, \rm K$ might be a limit seen in the extrasolar terrestrial planet population.
Does this interpretation still fit if we now see more sticking in hot collisions instead of smaller aggregates in cold collisions but with tempered dust?
We cannot give a confirmative answer in this paper, but two extremes seem possible. On one side, more sticking implies larger growth and these larger aggregates might seed planetesimal formation. This seems inconsistent with a $1000\, \rm K$ limit for planet formation. However, there are still many evolutionary steps from planetesimals to planets. On the other side, larger aggregates drift inward faster.  As the next thing inward is the sublimation line, the dust is taken out of the system. Therefore, the growth region might be drained of dust more readily, preventing the formation of planetesimals locally. This would be in agreement to fewer terrestrial exoplanets on orbits with higher temperatures. It has to be kept in mind though that temperatures also change locally over the lifetime of protoplanetary disks. What might be a preferential zone for growing dust at one time might be an inactive zone later or vice versa.
To answer, in what direction this might go, more elaborate simulations might be needed.
It is clear however that the onset of planet formation is different at high temperatures. 

\section{Conclusions}

In this work we present, for the first time, a new experimental setup that is capable of studying 
slow collisions of dust aggregates at high temperature. 
This does not allow for a wider interpretation on planet formation yet but we provide another piece in the preplanetary puzzle. We show that dust at high temperatures, starting at $900\, \rm K$ in our case aggregates differently, i.e., it likely grows much better as the sticking probability increases, which might also shift the bouncing barrier to larger aggregate sizes.

\begin{acknowledgements}
This project is supported by DFG grant WU 321/18-1. T.D. is funded by DLR (Deutsches Zentrum f\"{u}r Luft- und Raumfahrt) space administration with funds provided by the the BMWi (Bundesministerium f\"{u}r Wirtschaft und Energie) under grant 50 WM 1760.
\end{acknowledgements}


\begin{thebibliography}{40}	
	\bibitem[{Aumatell \& Wurm(2011)}]{Aumatell2011}
	Aumatell, G. \& Wurm, G. 2011, Monthly Notices of the Royal Astronomical
	Society: Letters, 418, L1
	
	\bibitem[{Aumatell \& Wurm(2013)}]{Aumatell2013}
	Aumatell, G. \& Wurm, G. 2013, Monthly Notices of the Royal Astronomical
	Society, 437, 690
	
	\bibitem[{Beitz {et~al.}(2011)Beitz, Güttler, Blum, Meisner, Teiser, \&
		Wurm}]{Beitz2011}
	Beitz, E., Güttler, C., Blum, J., {et~al.} 2011, The Astrophysical Journal,
	736, 34
	
	\bibitem[{Birnstiel {et~al.}(2010)Birnstiel, Dullemond, \&
		Brauer}]{Birnstiel2010}
	Birnstiel, T., Dullemond, C.~P., \& Brauer, F. 2010, A\&A, 513, A79
	
	\bibitem[{Birnstiel {et~al.}(2016)Birnstiel, Fang, \& Johansen}]{Birnstiel2016}
	Birnstiel, T., Fang, M., \& Johansen, A. 2016, Space Science Reviews, 205, 41
	
	\bibitem[{Blum {et~al.}(1998)Blum, Wurm, Poppe, \& Heim}]{Blum1998}
	Blum, J., Wurm, G., Poppe, T., \& Heim, L.-O. 1998, Earth, Moon, and Planets,
	80, 285
	
	\bibitem[{Bogdan {et~al.}(2019)Bogdan, Teiser, Fischer, Kruss, \&
		Wurm}]{Bogdan2019}
	Bogdan, T., Teiser, J., Fischer, N., Kruss, M., \& Wurm, G. 2019, Icarus, 319,
	133
	
	\bibitem[{de~Beule {et~al.}(2017)de~Beule, Landers, Salamon, Wende, \&
		Wurm}]{deBeule2017}
	de~Beule, C., Landers, J., Salamon, S., Wende, H., \& Wurm, G. 2017, ApJ, 837,
	59
	
	\bibitem[{Deckers \& Teiser(2014)}]{Deckers2014}
	Deckers, J. \& Teiser, J. 2014, The Astrophysical Journal, 796, 99
	
	\bibitem[{{Demirci} {et~al.}(2019){Demirci}, {Kruss}, {Teiser}, {Bogdan},
		{Jungmann}, {Schneider}, \& {Wurm}}]{Demirci2019}
	{Demirci}, T., {Kruss}, M., {Teiser}, J., {et~al.} 2019, \mnras, 484, 2779
	
	\bibitem[{Demirci {et~al.}(2017)Demirci, Teiser, Steinpilz, Landers, Salamon,
		Wende, \& Wurm}]{Demirci2017}
	Demirci, T., Teiser, J., Steinpilz, T., {et~al.} 2017, ApJ, 846, 48
	
	\bibitem[{{Dominik} \& {Tielens}(1997)}]{Dominik1997}
	{Dominik}, C. \& {Tielens}, A. G. G.~M. 1997, \apj, 480, 647
	
	\bibitem[{Drazkowska {et~al.}(2013)Drazkowska, Windmark, \&
		Dullemond}]{Drazkowska2013}
	Drazkowska, J., Windmark, F., \& Dullemond, C.~P. 2013, A\&A, 556, A37
	
	\bibitem[{Giordano {et~al.}(2008)Giordano, Russell, \& Dingwell}]{Giordano2008}
	Giordano, D., Russell, J.~K., \& Dingwell, D.~B. 2008, Earth and Planetary
	Science Letters, 271, 123
	
	\bibitem[{Golden(2013)}]{Golden2013}
	Golden, L.~M. 2013, Laboratory Experiments in Physics for Modern Astronomy -
	With Comprehensive Development of the Physical Principles (Springer)
	
	\bibitem[{Gundlach {et~al.}(2011)Gundlach, Kilias, Beitz, \&
		Blum}]{Gundlach2011}
	Gundlach, B., Kilias, S., Beitz, E., \& Blum, J. 2011, Icarus, 214, 717
	
	\bibitem[{{Gundlach} {et~al.}(2018){Gundlach}, {Schmidt}, {Kreuzig},
		{Bischoff}, {Rezaei}, {Kothe}, {Blum}, {Grzesik}, \& {Stoll}}]{Gundlach2018}
	{Gundlach}, B., {Schmidt}, K.~P., {Kreuzig}, C., {et~al.} 2018, \mnras, 479,
	1273
	
	\bibitem[{G\"uttler {et~al.}(2010)G\"uttler, Blum, Zsom, Ormel, \&
		Dullemond}]{Guettler2010}
	G\"uttler, C., Blum, J., Zsom, A., Ormel, C.~W., \& Dullemond, C.~P. 2010,
	A\&A, 513, A56
	
	\bibitem[{{Heim} {et~al.}(1999){Heim}, {Blum}, {Preuss}, \& {Butt}}]{Heim1999}
	{Heim}, L.-O., {Blum}, J., {Preuss}, M., \& {Butt}, H.-J. 1999, Physical Review
	Letters, 83, 3328
	
	\bibitem[{Hubbard(2015)}]{Hubbard2015}
	Hubbard, A. 2015, Icarus, 254
	
	\bibitem[{Kataoka {et~al.}(2013)Kataoka, Tanaka, Okuzumi, \&
		Wada}]{Kataoka2013}
	Kataoka, A., Tanaka, H., Okuzumi, S., \& Wada, K. 2013, A\&A, 557, L4
	
	\bibitem[{Kelling \& Wurm(2009)}]{Kelling2009}
	Kelling, T. \& Wurm, G. 2009, Phys. Rev. Lett., 103, 215502
	
	\bibitem[{Kelling {et~al.}(2014)Kelling, Wurm, \& Köster}]{Kelling2014}
	Kelling, T., Wurm, G., \& Köster, M. 2014, The Astrophysical Journal, 783, 111
	
	\bibitem[{Keppler {et~al.}(2018)Keppler, {Benisty, M.}, {M\"uller, A.},
		{Henning, Th.}, {van Boekel, R.}, {Cantalloube, F.}, {Ginski, C.}, {van
			Holstein, R. G.}, {Maire, A.-L.}, {Pohl, A.}, {Samland, M.}, {Avenhaus, H.},
		{Baudino, J.-L.}, {Boccaletti, A.}, {de Boer, J.}, {Bonnefoy, M.}, {Chauvin,
			G.}, {Desidera, S.}, {Langlois, M.}, {Lazzoni, C.}, {Marleau, G.-D.},
		{Mordasini, C.}, {Pawellek, N.}, {Stolker, T.}, {Vigan, A.}, {Zurlo, A.},
		{Birnstiel, T.}, {Brandner, W.}, {Feldt, M.}, {Flock, M.}, {Girard, J.},
		{Gratton, R.}, {Hagelberg, J.}, {Isella, A.}, {Janson, M.}, {Juhasz, A.},
		{Kemmer, J.}, {Kral, Q.}, {Lagrange, A.-M.}, {Launhardt, R.}, {Matter, A.},
		{M\'enard, F.}, {Milli, J.}, {Molli\`ere, P.}, {Olofsson, J.}, {P\'erez, L.},
		{Pinilla, P.}, {Pinte, C.}, {Quanz, S. P.}, {Schmidt, T.}, {Udry, S.},
		{Wahhaj, Z.}, {Williams, J. P.}, {Buenzli, E.}, {Cudel, M.}, {Dominik, C.},
		{Galicher, R.}, {Kasper, M.}, {Lannier, J.}, {Mesa, D.}, {Mouillet, D.},
		{Peretti, S.}, {Perrot, C.}, {Salter, G.}, {Sissa, E.}, {Wildi, F.}, {Abe,
			L.}, {Antichi, J.}, {Augereau, J.-C.}, {Baruffolo, A.}, {Baudoz, P.},
		{Bazzon, A.}, {Beuzit, J.-L.}, {Blanchard, P.}, {Brems, S. S.}, {Buey, T.},
		{De Caprio, V.}, {Carbillet, M.}, {Carle, M.}, {Cascone, E.}, {Cheetham, A.},
		{Claudi, R.}, {Costille, A.}, {Delboulb\'e, A.}, {Dohlen, K.}, {Fantinel,
			D.}, {Feautrier, P.}, {Fusco, T.}, {Giro, E.}, {Gluck, L.}, {Gry, C.},
		{Hubin, N.}, {Hugot, E.}, {Jaquet, M.}, {Le Mignant, D.}, {Llored, M.},
		{Madec, F.}, {Magnard, Y.}, {Martinez, P.}, {Maurel, D.}, {Meyer, M.},
		{M\"oller-Nilsson, O.}, {Moulin, T.}, {Mugnier, L.}, {Orign\'e, A.}, {Pavlov,
			A.}, {Perret, D.}, {Petit, C.}, {Pragt, J.}, {Puget, P.}, {Rabou, P.},
		{Ramos, J.}, {Rigal, F.}, {Rochat, S.}, {Roelfsema, R.}, {Rousset, G.},
		{Roux, A.}, {Salasnich, B.}, {Sauvage, J.-F.}, {Sevin, A.}, {Soenke, C.},
		{Stadler, E.}, {Suarez, M.}, {Turatto, M.}, \& {Weber, L.}}]{Keppler2018}
	Keppler, M., {Benisty, M.}, {M\"uller, A.}, {et~al.} 2018, A\&A, 617, A44
	
	\bibitem[{{Kimura} {et~al.}(2015){Kimura}, {Wada}, {Senshu}, \&
		{Kobayashi}}]{Kimura2015}
	{Kimura}, H., {Wada}, K., {Senshu}, H., \& {Kobayashi}, H. 2015, \apj, 812, 67
	
	\bibitem[{Klahr {et~al.}(2018)Klahr, Pfeil, \& Schreiber}]{Klahr2018}
	Klahr, H., Pfeil, T., \& Schreiber, A. 2018, Instabilities and Flow Structures
	in Protoplanetary Disks: Setting the Stage for Planetesimal Formation, ed.
	H.~J. Deeg \& J.~A. Belmonte (Cham: Springer International Publishing),
	2251--2286
	
	\bibitem[{Kruss {et~al.}(2016)Kruss, Demirci, Koester, Kelling, \&
		Wurm}]{Kruss2016}
	Kruss, M., Demirci, T., Koester, M., Kelling, T., \& Wurm, G. 2016, The
	Astrophysical Journal, 827, 110
	
	\bibitem[{Kruss {et~al.}(2017)Kruss, Teiser, \& Wurm}]{Kruss2017}
	Kruss, M., Teiser, J., \& Wurm, G. 2017, A\&A, 600, A103
	
	\bibitem[{Kruss \& Wurm(2018)}]{Kruss2018}
	Kruss, M. \& Wurm, G. 2018, The Astrophysical Journal, 869, 45
	
	\bibitem[{Musiolik {et~al.}(2016)Musiolik, Teiser, Jankowski, \&
		Wurm}]{Musiolik2016}
	Musiolik, G., Teiser, J., Jankowski, T., \& Wurm, G. 2016, The Astrophysical
	Journal, 818, 16
	
	\bibitem[{Musiolik \& Wurm(2019)}]{Musiolik2019}
	Musiolik, G. \& Wurm, G. 2019, The Astrophysical Journal, 873, 58
	
	\bibitem[{Okuzumi {et~al.}(2012)Okuzumi, Tanaka, Kobayashi, \&
		Wada}]{Okuzumi2012}
	Okuzumi, S., Tanaka, H., Kobayashi, H., \& Wada, K. 2012, ApJ, 752, 106
	
	\bibitem[{Saito \& Sirono(2011)}]{Saito2011}
	Saito, E. \& Sirono, S.-i. 2011, The Astrophysical Journal, 728, 20
	
	\bibitem[{Steinpilz {et~al.}(2019)Steinpilz, Teiser, \& Wurm}]{Steinpilz2019}
	Steinpilz, T., Teiser, J., \& Wurm, G. 2019, The Astrophysical Journal, 874, 60
	
	\bibitem[{Teiser \& Wurm(2009)}]{Teiser2009}
	Teiser, J. \& Wurm, G. 2009, Monthly Notices of the Royal Astronomical Society,
	393, 1584
	
	\bibitem[{Weidenschilling(1977)}]{Weidenschilling1977}
	Weidenschilling, S.~J. 1977, MNRAS, 180, 57
	
	\bibitem[{Windmark {et~al.}(2012)Windmark, Birnstiel, G\"uttler, Blum,
		Dullemond, \& Henning}]{Windmark2012}
	Windmark, F., Birnstiel, T., G\"uttler, C., {et~al.} 2012, A\&A, 540, A73
	
	\bibitem[{Wurm {et~al.}(2005)Wurm, Paraskov, \& Krauss}]{Wurm2005}
	Wurm, G., Paraskov, G., \& Krauss, O. 2005, Icarus, 178, 253
	
	\bibitem[{Yang {et~al.}(2017)Yang, Johansen, \& Carrera}]{Yang2017}
	Yang, C.-C., Johansen, A., \& Carrera, D. 2017, A\&A, 606, A80
	
	\bibitem[{Zsom {et~al.}(2010)Zsom, Ormel, G\"uttler, Blum, \&
		Dullemond}]{Zsom2010}
	Zsom, A., Ormel, C.~W., G\"uttler, C., Blum, J., \& Dullemond, C.~P. 2010,
	A\&A, 513, A57
\end{thebibliography}
\end{document}